# Large gender and age differences in hand disinfection behavior during the COVID-19 pandemic: Field data from Swiss retail stores

Frauke von Bieberstein[1], Anna-Corinna Kulle[1], Stefanie Schumacher*[1]

[1] University of Bern

March 2022

## Abstract

Hand hygiene is one of the key low-cost measures proposed by the World Health Organization (WHO) to contain the spread of COVID-19. In a field study conducted during the pandemic in June and July 2020 in Switzerland, we captured the hand disinfection behavior of customers in five stores of a large retail chain (n = 8,245). The study reveals considerable differences with respect to gender and age: Women were 8.3 percentage points more likely to disinfect their hands compared to men. With respect to age, we identified a steep increase across age groups, with people age 60 years and older disinfecting their hands significantly more often than younger adults (>+16.7 percentage points) and youth (>+ 31.7 percentage points). A validation study conducted in December 2020 (n = 1,918) confirmed the gender and age differences at a later point in the pandemic. In sum, the differences between gender and age groups are substantial and should be considered in the design of protective measures to ensure clean and safe hands.

---

*corresponding author: stefanie.schumacher@unibe.ch

# Introduction

There have been over 470 million confirmed cases of and more than 6 million deaths globally from the novel coronavirus (COVID-19) as of March 28, 2022[1]. The WHO recommends, among other procedures such as wearing a properly fitting mask, hand hygiene with alcohol-based hand rub or soap and water as an effective low-cost measure for the public to reduce the spread of SARS-CoV-2[2]. Consequently, hand hygiene recommendations have been issued by governments around the world to fight the pandemic[3–6]. However, survey evidence suggests substantial differences in compliance with hand hygiene guidance depending on gender and age[7,8]. These demographic factors that influence compliance are important in designing suitable campaigns to increase hand hygiene.

Thus far, we lack knowledge of actual hand hygiene behavior for different age groups and genders during the pandemic. However, there has been substantial survey research on self-reports of hand hygiene during the COVID-19 crisis[7–10]. This research can provide important indications regarding the influence of these demographic factors. Two large representative surveys conducted in the United States found that men and younger people reported less frequent hand hygiene during the COVID-19 pandemic compared to women and older people[7,8]. This gender difference is mostly in line with survey evidence from other countries[9,10]. Older age seems to be associated with higher intended compliance[11]. However, researchers in Saudi Arabia found that hand hygiene knowledge tested during the COVID-19 pandemic was greater for younger age groups than for older age groups[12].

Although surveys represent an important way to advance knowledge, research in fields such as patient behavior suggests that respondents tend to overestimate their own readiness to perform (socially) desirable behaviors[13,14]. Recent studies on compliance in the context of COVID-19 have indicated a similar pattern[15,16]. Therefore, it is important to complement survey research with studies that observe actual behavior. Two studies on actual hand hygiene behavior based on data from U.S. hospitals showed that hand hygiene first increased substantially during the pandemic and later decreased[17,18]. However, as these studies used automated measurements by disinfection dispensers, they could not report on gender and age. A study in Germany that investigated disinfection behavior by healthcare workers also showed an increase in hand hygiene during the pandemic[19]. In this study, most participants were female, and gender and age differences were not reported. A smaller-scale study in a hospital and mall in Nigeria showed that most visitors did not wash or sanitize their hands[20]. There were no gender differences, and the authors noted that age differences were likely due to the age differences between the two locations.

The present study addresses the gap in knowledge of actual hand hygiene behavior during COVID-19 for different age groups and genders. We present the results of a field study conducted in June and July 2020 in five stores of a large Swiss retail chain. Unobtrusive observers captured the actual hand disinfection behavior of customers entering the store, including estimates of the customer's gender and age group. Furthermore, we present the results of a validation study conducted in December 2020 in one of the five stores.

# Methods

**Study setting and design.** Following new government regulations based on WHO recommendations, hand disinfection dispensers were installed in the entrance areas of supermarkets and all other stores in Switzerland in March 2020. At the time of the study, customers visiting a supermarket were encouraged to disinfect their hands through a poster by the Swiss Ministry of Health, recommending increased hand hygiene as one measure to contain the crisis. We collected data on customers' hand hygiene behavior in five Swiss supermarkets over five days in June and July 2020. At the point in time of the measurement, the pandemic was in a relatively mild state, with most lock-down restrictions in Switzerland loosened and generally low infection and death numbers[21]. In each store, data collection occurred on at least two independent days. Data from three of the stores were used as a control group for an experimental study that focused on increasing hand disinfection behavior[22]. The complete dataset presented here, as well as the validation data discussed below, is unique to this study. The ethical standards for the data collection and the experimental study were approved by the Faculty of Business Administration, Economics and Social Sciences of the University of Bern (June 3, 2020 – serial number 072020).

**On-site data collection.** Unobtrusive incognito observers captured hand disinfection behavior in the five supermarkets. The stores for this study were, among other criteria, selected so that there was a spot for unnoticed unobtrusive observation, only one entrance to observe, and a maximum average number of customers for any given point in time so that the observers would be able to capture all incoming customers. The observer manually captured hand disinfection behavior on a paper sheet for each hour, indicating the incoming person's behavior (disinfecting hands, yes or no) and estimating the person's gender and age. Gender was captured as binary (female or male) and age as categorical, with four distinct age groups. The four age groups captured were youth (12–17 years), adult (18–59 years), golden (60–74 years), and old (75 years and older). In addition to these two main demographic characteristics, the day of the week and the store were recorded as control variables. The observation data captured on paper were transferred to an Excel spreadsheet. The correct transfer was checked by three people independently.

**Validation study.** To ensure the validity of the study's findings over time, data were collected a second time, in December 2020 (approximately six months after the original data collection) on one day in one of the five stores in the study sample. At that point in time, Switzerland was in a more severe pandemic state compared to the summer, with lock-down measures in place and a high and increasing number of daily COVID-19 infections and related deaths[21,23].

**Statistical analyses.** All analyses were performed using Stata 17 software (StataCorp Inc.). As all variables in this study were categorical, differences between the variables were investigated using two-sided chi-square tests. In a multiple regression analysis, hand disinfection behavior was regressed on age group, gender, day, and store. The same analysis was run for each gender. We conducted a separate analysis for the observations of the validation study, regressing disinfection behavior on age group and gender. The results are robust to the inclusion of the additional data from December in the main data set. As the outcome variable is bounded [0,1], we report results from logistic regression analysis.

## Results

**Descriptive statistics.** In total, 8,245 customers were observed across the five field days of the main study. Of these customers, 62.8% were classified as female. Regarding age, 63.6% of observations were classified in the age group adult, 28.5% in the age group golden, 4.7% in the age group youth, and 3.1% in the age group old. The female share was similar in the age groups youth and adult with 60.9% each, and somewhat higher in the age groups golden and old with 66.7% and 69.5%, respectively.

**Non-parametric analyses.** The hand disinfection rate across the sample was 55.6%. There were statistically significant differences in hand disinfection behavior regarding the gender and age groups. As shown in Figure 1, the overall hand disinfection rate for women is +8.3 percentage points higher than that for men ($p < 0.001$, two-sided chi-square test).

*Figure 1:* Hand disinfection rates by gender.

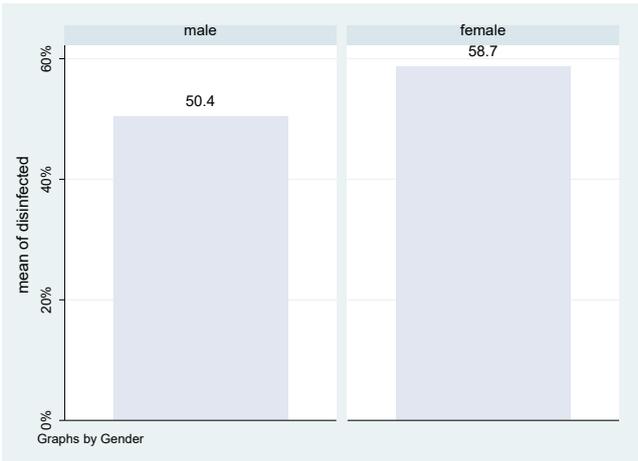

With regard to age, we observed substantial differences in hand disinfection behavior between the four groups. As Figure 2 shows, people in the two oldest age groups disinfected their hands significantly more often than younger adults (+18.5 percentage points for the golden age group; +16.7 percentage points for the old age group, $p < 0.001$ two-sided chi-square tests) and youth (+33.5 percentage points for the golden age group; +31.7 percentage points for the old age group, $p < 0.001$ two-sided chi-square tests), revealing a steep increase from youth to older adults. Pairwise comparisons between the age groups revealed that all but the difference between the golden and old age groups was highly statistically significant ($p < 0.001$, two-sided chi-square tests).

*Figure 2:* Hand disinfection rates by age group.

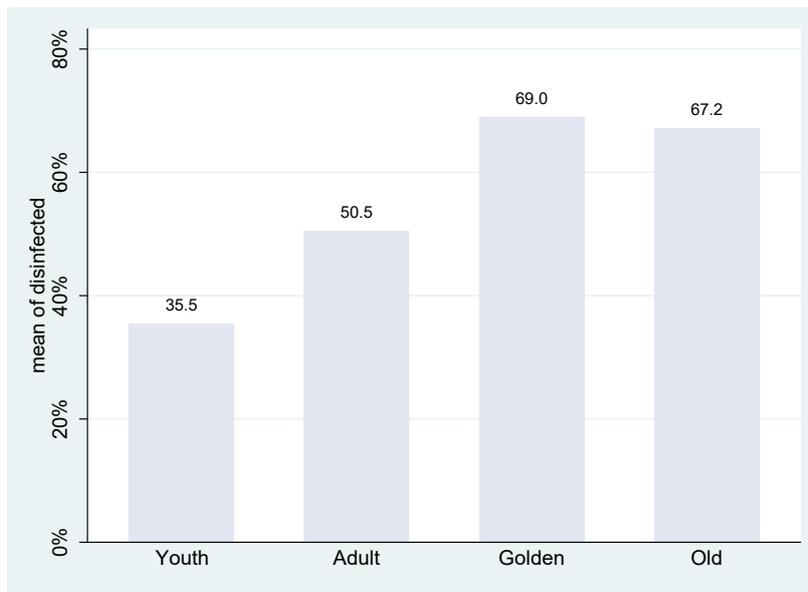

Assessing the gender difference for each age group, Table 1 shows that the gender difference in hand disinfection behavior is greatest and highly statistically significant, with +9.4 percentage points (p < 0.001, two-sided chi-square test) in the largest age group adult. The gender differences in the age groups golden and old have the same direction, but are only statistically weakly significant for the golden age group (statistically significant at the < 5% level if observations from December are included), and not significant for the age group old. There is no observable difference between the two genders in the age group youth.

*Table 1:* Overview of hand disinfection rates by age group and gender.

| | Total (n = 8,245) | Age youth (1) (n = 389) | Age adult (2) (n = 5,246) | Age golden (3) (n = 2,351) | Age old (4) (n = 259) | p-value (1) vs. (2) | p-value (1) vs. (3) | p-value (1) vs. (4) | p-value (2) vs. (3) | p-value (2) vs. (4) | p-value (3) vs. (4) |
|---|---|---|---|---|---|---|---|---|---|---|---|
| Total | 55.6 | 35.5 | 50.5 | 69.0 | 67.2 | <0.000 | <0.000 | <0.000 | <0.000 | <0.000 | 0.541 |
| Female (n=5,179) | 58.7 | 35.4 | 54.2 | 70.2 | 69.4 | <0.000 | <0.000 | <0.000 | <0.000 | <0.000 | 0.840 |
| Male (n=3,066) | 50.4 | 35.5 | 44.8 | 66.8 | 62.0 | 0.026 | <0.000 | <0.000 | <0.000 | 0.003 | 0.397 |

Notes: The table shows p-values from two-sided chi-square tests.

**Multiple regression analysis.** Results of multiple logistic regressions for predicting hand disinfection behavior are shown in Table 2. All models show a statistically highly significant difference among the age groups adult, golden, and old age groups versus the youth age group, as well as a statistically highly significant gender difference. The magnitude of the age group and gender differences remained stable when controlling for weekday (Model 3) and store (Model 4). The magnitude and significance levels also remained stable when additional controls for date, time, and observer were included and when the December validation study data were included in the dataset. Multiple regression analysis run separately for both genders showed that the age group differences were stable and statistically highly significant for men (coefficient adults, 0.560; golden, 1.397; and old, 1.209) and women (coefficient adults, 1.103; golden, 1.774; and old, 1.874).

*Table 2:* Hand disinfection behavior during the study (June/July 2020).

|  | Disinfected (1) | Disinfected (2) | Disinfected (3) | Disinfected (4) |
|---|---|---|---|---|
| Adult | 0.618*** | 0.621*** | 0.628*** | 0.890*** |
|  | (0.110) | (0.110) | (0.110) | (0.115) |
| Golden | 1.400*** | 1.389*** | 1.395*** | 1.616*** |
|  | (0.115) | (0.116) | (0.116) | (0.121) |
| Old | 1.315*** | 1.295*** | 1.287*** | 1.618*** |
|  | (0.170) | (0.170) | (0.170) | (0.176) |
| Female |  | 0.302*** | 0.304*** | 0.350*** |
|  |  | (0.047) | (0.047) | (0.048) |
| Thursday |  |  | -0.070 | 0.036 |
|  |  |  | (0.052) | (0.054) |
| Friday |  |  | 0.104* | 0.070 |
|  |  |  | (0.061) | (0.068) |
| Store_2 |  |  |  | -0.730*** |
|  |  |  |  | (0.077) |
| Store_3 |  |  |  | 0.461*** |
|  |  |  |  | (0.085) |
| Store_4 |  |  |  | 0.131* |
|  |  |  |  | (0.077) |
| Store_5 |  |  |  | -0.344*** |
|  |  |  |  | (0.074) |
| Constant | -0.598*** | -0.786*** | -0.788*** | -0.929*** |
|  | (0.106) | (0.111) | (0.115) | (0.132) |
| Observations | 8,245 | 8,245 | 8,245 | 8,245 |

The table presents estimates of logistic regressions. Standard errors are in parentheses. Significance levels:
* $p < 0.10$, ** $p < 0.05$, *** $p < 0.01$

**Validation study.** To examine prevailing age and gender differences at a later point in the pandemic, we gathered additional data in December 2020. The age differences were even more pronounced when the winter validation data were compared to the summer data only from the store of the validation study (see Table 3). In addition to these continued behavioral age and gender differences, the validation study showed that overall hand disinfection behavior may have dropped over the six-month period. We observed an average hand disinfection rate of 44.4% in winter 2020 compared to 55.6% across stores and 61.9% in the store of the validation study in summer 2020. However, the reduction in hand disinfection rates over time should be regarded as a preliminary insight that needs further verification, as in the validation study, behavior was observed only in one store on one day.

*Table 3:* Comparison of hand disinfection behavior in summer (June/July) and winter (December) 2020.

|  | Summer, all stores (1) (n = 8,245) | Summer, Store_4 (2) (n = 1,591) | Winter, Store_4 (3) (n = 1,918) | p-value (1) vs. (2) | p-value (2) vs. (3) |
|---|---|---|---|---|---|
| Total | 55.6 | 61.9 | 44.4 | <0.001 | <0.001 |
| Female | 58.7 | 66.3 | 47.2 | <0.001 | <0.001 |
| Male | 50.4 | 55.0 | 39.9 | <0.001 | <0.001 |
| Age youth | 35.5 | 34.3 | 16.7 | <0.001 | 0.002 |
| Age adult | 50.5 | 61.5 | 39.1 | <0.001 | <0.001 |
| Age golden | 69.0 | 69.1 | 57.1 | <0.001 | <0.001 |
| Age old | 67.2 | 60.9 | 65.4 | 0.74 | 0.685 |

Notes: The table shows p-values from two-sided chi-square tests.

The results were confirmed with logistic regression analyses. Two regression models were estimated, using the additional data collected in winter 2020 (see Table 4). The differences in hand disinfection behavior between age groups (Model 1) and between genders (Model 2) remained statistically highly significant. Further, the age group differences between the adult and youth age groups and the adult and old age groups increased.

*Table 4:* Hand disinfection behavior during the validation study (December 2020).

|  | Disinfected (1) | Disinfected (2) |
|---|---|---|
| Adult | 1.165*** | 1.156*** |
|  | (0.247) | (0.249) |
| Golden | 1.894*** | 1.906*** |
|  | (0.253) | (0.256) |
| Old | 2.244*** | 2.307*** |
|  | (0.318) | (0.322) |
| Female |  | 0.368*** |
|  |  | (0.099) |
| Constant | -1.609*** | -1.842*** |
|  | (0.239) | (0.252) |
| Observations | 1,918 | 1,918 |

The table presents estimates of logistic regressions. Standard errors are in parentheses. Significance levels: * $p < 0.10$, ** $p < 0.05$, *** $p < 0.01$

# Discussion

Hand hygiene is a key element in most national COVID-19 prevention programs and continues to be very important in preventing the spread of other infectious diseases[24,25]. To the best of our knowledge, this is the first observational study to capture actual hand hygiene behavior with a focus on gender and age groups during the COVID-19 pandemic. We found considerable differences in the observed behavior.

For women, we found a statistically significantly higher rate of hand disinfection behavior of +8.3 percentage points compared to men. This difference in actual behavior is in the same direction, but even higher compared to gender differences identified in surveys from the United States. There, +5.6 percentage points more women stated that they often or always use hand sanitizers compared to men[7] and +5.3 percentage points more women stated that they wash their hands before eating at a restaurant compared to men[8]. The present findings are also in line with COVID-19 studies assessing gender differences in beliefs and behaviors beyond hand hygiene. For example, a survey study across eight Organization for Economic Co-operation and Development countries found that women are more likely to perceive COVID-19 as a very serious health problem, to agree with protective public policy measures that restrict behavior, and to comply with them[26]. Relatedly, people who were more concerned about their personal risk of contracting SARS-CoV-2 reported more frequent handwashing and hand sanitizing[7].

With regard to age, we observed a steep increase in the disinfection rate from youth (12–17 years) to younger adults to people age 60 years and older. People age 60 years and older disinfect their hands significantly more often than youth (>+ 31.7 percentage points) and younger adults (>+16.7 percentage points). One potential reason for the more compliant behavior in the age groups 60 years and older might be an accurately higher perceived risk of contracting a severe case of COVID-19, with most COVID-19-related hospitalizations and fatalities in these age groups[27,28].

Although survey evidence has also shown an increase in hand hygiene with age, the differences were much smaller among older adults. For instance, previous survey studies showed that the stated use of hand sanitizers differed only by at most 1.5 percentage points between the age groups 25–44, 45–65, and older, while there was a substantial difference of 12 percentage points for the youngest age group[7]. A study on handwashing before eating in a restaurant also showed a positive association with age, with the highest difference of 9.9 percentage points between age groups[8]. Although the present study was conducted at a different point in time in the pandemic in a different country and setting, it still highlights the importance of complementing survey research with observational studies capturing actual behavior. The validation study conducted six months later identified similar age and gender differences as the main study at a later point in the pandemic. In addition, although the dataset is considerably smaller, it provides some indication that overall hand disinfection behavior declined for both genders and in all age groups, except the age group 75 years and older, over the course of the pandemic, although the state of the pandemic was more severe in winter 2020 than in the summer of the main study. Thus, it can be reasoned that fatigue in complying with protective measures such as hand hygiene developed over time. This appeared not to be the case only for the age group 75 years and older, potentially because their individual perceived need for protection continued to be high, due to much higher hospitalization and fatality rates than in the other age groups[27,28].

When considering the joint effect of gender and age, the gender difference is statistically significant in the adult and golden age groups and shows the same direction for the old age group. Similar differences can be found in the validation study six months later. Interestingly, even in the most vulnerable age groups 60 years and older, a gender difference can be observed. In contrast, we did not find a gender difference in the youngest group of people aged 12–17 years.

Overall, the results for gender and age differences are in line with those of other studies on compliance with other COVID-19 protective measures, such as social distancing and mask wearing, which also showed that being female and older predict compliance[29–32].

The main limitation of the present study is that all data were recorded manually, which makes the process more error-prone than the automatic measurement of hand disinfection. Manual recording was chosen as this was the only option available at the time of the study that allowed for capturing estimates for gender and age. To keep the error rate as low as possible, only two genders and four fairly well distinguishable age groups were chosen. To keep the recording simple and accurate, the observers did not distinguish between customers entering alone and in groups. Each incoming customer was recorded as an independent observation; therefore, we cannot assess potential group behavior. Moreover, the study was limited to five stores in the German-speaking region of Switzerland, and there may be geographic differences that cannot be accounted for.

The study findings open up several avenues for future research. First, further research would benefit from testing whether the gender and age differences observed in this setting are present in other settings and/or other cultural and geographic environments. Second, future studies could explore the specific mechanisms behind the observed differences. Third, as the findings suggest that compliance with hand hygiene can be improved most in younger age groups and in men, it is important to test how communication can be tailored to increase compliant behavior in these groups. In conclusion, these findings help understand how hand hygiene compliance differs across gender and age group and thus can help to better design and communicate hand hygiene recommendations.

**Acknowledgements**

We thank our partner for granting access to the stores and for enabling us to conduct this study. We also thank Lorenz Affolter, Daniel Frey, Nicolas Hafner, Greta Ingendaay, Saskia Oetterli, Aarusza Ramachandran, Laura Sennhauser, and Fredrik Sitje for collecting the data and providing excellent research assistance.

**Author contributions**

FvB, ACK, and SS wrote the main manuscript text. FvB, ACK, and SS conducted and reviewed the statistical analyses; ACK and SS prepared the tables and figures. FvB, ACK, and SS planed and supervised the study. ACK supervised the data collection. FvB, ACK, and SS reviewed and approved the final draft of the manuscript.